\documentclass[11pt,a4paper]{article}
\usepackage[a4paper,total={160mm,220mm}]{geometry}
\usepackage{amsmath,amssymb}
\usepackage{cite}
\usepackage[colorlinks=true,allcolors=blue,linktocpage=true]{hyperref}
\usepackage[capitalise]{cleveref}
\usepackage{caption}
\usepackage{subcaption}
\usepackage{graphicx}
\graphicspath{{figures/}}
\usepackage{xcolor}
\usepackage{tensor}
\usepackage{booktabs}

\newcommand{\abs}[1]{\left\lvert{#1}\right\rvert}
\newcommand{\absl}[1]{\lvert{#1}\rvert}
\newcommand{\md}{\mathrm{d}}
\newcommand{\me}{\mathrm{e}}

\renewcommand{\vec}[1]{\mathbf{#1}}

\newcommand{\jeff}{\vec{j}_{\mathrm{eff}}}
\newcommand{\vcav}{V_{\mathrm{cav}}}
\newcommand{\Jhat}{\hat{J}}
\newcommand{\nmax}{N_{\mathrm{max}}}
\newcommand{\wtot}{W_{\mathrm{tot}}}

\begin{document}

\numberwithin{equation}{section}
\title{
\vspace*{-0.5cm}
{\scriptsize \mbox{}\hfill MITP-25-083}\\
\vspace{3.5cm}
\Large{\textbf{Signatures of High-Frequency Gravitational Waves in Electromagnetic Cavities}}
\vspace{0.5cm}
}

\author{Sebastian Schenk,$^{1,2}$ Kristof Schmieden,$^{1,3}$ and Pedro Schwaller$^1$
\\[2ex]
\small{\em $^{1}$PRISMA$^+$ Cluster of Excellence \& Mainz Institute for Theoretical Physics,} \\
\small{\em Johannes Gutenberg-Universit\"at Mainz, 55099 Mainz, Germany}
\\[0.5ex]
\small{\em $^{2}$Institute for Astroparticle Physics, Karlsruhe Institute of Technology,} \\
\small{\em Hermann-von-Helmholtz-Platz 1, 76344 Eggenstein-Leopoldshafen, Germany}
\\[0.5ex]
\small{\em $^{3}$Physikalisches Institut, Rheinische Friedrich-Wilhelms-Universität Bonn,} \\
\small{\em 53115 Bonn, Germany}
\\[0.8ex]}
\date{}
\maketitle

\begin{abstract}
\noindent
Similar to axions, gravitational waves (GW) can induce oscillating electromagnetic fields inside electromagnetic cavities.
We explore their experimental sensitivity to monochromatic and non-monochromatic GW signals, using the total deposited energy as a primary measure.
Focusing on cylindrical and spherical cavities, we present the coupling coefficients of GWs to the dominant electromagnetic resonances in transverse-traceless gauge, which is most appropriate in this regime.
By considering the superposition of degenerate modes, we further examine their angular sensitivity.
In addition, we calculate the response of a spherical cavity to non-monochromatic GWs emitted by primordial black hole mergers.
We find that, for transient signals, a high quality factor with $Q \gtrsim 10^5$ does not necessarily enhance experimental sensitivity.
In fact, even in the most optimistic scenario, only mergers within the solar system yield an observable energy deposit in the cavity.
\end{abstract}

\newpage

\tableofcontents

\section{Introduction}
\label{sec:introduction}

Gravitational waves (GWs) at GHz frequencies and beyond naturally probe the characteristic size of laboratory and tabletop experiments on Earth.
Over the past few years, a myriad of detection strategies has been proposed~\cite{Aggarwal:2025noe}, including the mechanical response to GWs~\cite{Arvanitaki:2012cn, Goryachev:2014yra, Aggarwal:2020umq, Berlin:2023grv, Kahn:2023mrj, Domcke:2024mfu, Carney:2024zzk, Schnabel:2024hem}, interferometric methods~\cite{Bringmann:2023gba, Heisig:2025oim} (see also~\cite{Domcke:2024abs}), or their coupling to spin systems~\cite{Ito:2019wcb, Ito:2022rxn, Liang:2025vfv}.
Here, we focus on their detection based on the inverse Gertsensthein effect, which describes the conversion of gravitational into electromagnetic waves in the vicinity of electromagnetic background fields~\cite{Gertsenshtein:1962}.
The underlying principle is identical to the case of axions, which may similarly be converted into photons in the presence of strong magnetic fields~\cite{Raffelt:1987im}.
These photons can accumulate in electromagnetic cavities, such that a sizable electromagnetic signal develops if the characteristic frequency of the signal corresponds to a resonant frequency of the cavity.
This detection strategy and required technology is broadly employed in axion searches, and there is now an extensive effort dedicated to both exploiting and improving their sensitivity to incorporate GW searches (see, e.g., \cite{Ballantini:2003nt, Ejlli:2019bqj, Berlin:2021txa, Domcke:2022rgu, Domcke:2023bat, Gatti:2024mde, Valero:2024ncz, Domcke:2024eti, Capdevilla:2024cby, Capdevilla:2025omb,Pappas:2025zld,Kim:2025izt}), or even exploring distinct proposals to target high-frequency GWs~\cite{Ballantini:2005am, Gao:2023gph, Schmieden:2023fzn, Alesini:2023qed, Navarro:2023eii, Garcia-Cely:2025mgu, Kharzeev:2025lyu, Takai:2025cyy}.

While ground-based laser interferometer~\cite{LIGOScientific:2016aoc} and pulsar timing array~\cite{NANOGrav:2023gor, EPTA:2023fyk, Reardon:2023gzh, Xu:2023wog} measurements have probed the dynamics of black hole mergers at low to intermediate frequencies, there are no confirmed astrophysical mechanisms to produce GWs at GHz frequencies or above.
Consequently, the observation of a high-frequency GW would provide compelling evidence for exotic sources, such as primordial black holes (PBHs), boson stars, or even new physics beyond the Standard Model in the early Universe~\cite{Aggarwal:2025noe}.

Sensitivity estimates for cavity experiments aimed at detecting high-frequency GW signals are primarily governed by the total energy that is deposited inside the cavity as the GW passes through the apparatus.
Close to the resonant frequencies of the cavity, this energy is mainly stored in the amplified oscillations of the electromagnetic field, which dominate over the mechanical excitations of the detector itself.
In this case, the sensitivity can be characterized by a dimensionless coupling coefficient, which captures the overlap between the effective current induced by the incoming GW and the resonant modes of the cavity, while neglecting the detector's motion.
In the regime where the incoming GW probes the characteristic length scale of the detector, $\omega L \gtrsim 1$, this approximation, however, is only viable in transverse-traceless (TT) gauge.
In other coordinate frames, this motion may become significant and must be carefully taken into account~\cite{Ratzinger:2024spd}. 
In this work, we explore the experimental sensitivity of both a cylindrical and a spherical cavity for GW searches at frequencies near electromagnetic resonances. We present the coupling coefficients for the dominant resonances in TT gauge, which provide an accurate approximation to the exact signal in this scenario, and obtain the sensitivity of three benchmark setups to monochromatic GWs produced by axion superradiance around PBHs.

The overlap function is only an accurate representation of the experimental sensitivity for monochromatic GW signals, while the energy deposited by non-monochromatic GW is instead limited by the time that the signal spends inside the cavity's resonance bands.
A prime example of a  non-monochromatic source at GHz frequencies are PBH binary mergers. 
During their inspiral phase, at low frequencies, the GW frequency increases with time.
This transient GW signal deposits energy in the cavity only when its oscillations are in phase with the eigenmodes of the cavity's resonance bands.
Rather than being limited by the damping time, characterized by the cavity's quality factor, the resonant amplification of photons is constrained by the time scale before decoherence of these oscillations occurs, significantly impacting the total energy deposited.
Here, we account for these aspects to estimate the total energy deposited inside the cavity and assess the experimental sensitivity for transient signals.
When applied to GWs emitted from PBH mergers, our results suggest that a high quality factor with $Q \gtrsim 10^5$ rarely enhances the experimental sensitivity.
Even under the most favorable conditions, only PBH mergers occurring within the solar system deposit an observable amount of energy inside the cavity, posing a significant challenge for experimental searches.

Our work is structured as follows.
In \cref{sec:gwDetectionCavities}, we briefly review the dynamics underlying the conversion of gravitational into electromagnetic waves.
We pay special attention to suitable coordinates frames where, close to the resonant frequencies, these dynamics reduce to overlap functions between the effective current induced by the incoming GW and the resonant excitations of the cavity.
We present these overlap functions both for cylindrical and spherical cavities in \cref{sec:couplingCoefficients}, and briefly discuss the dominant couplings.
In \cref{sec:beyondMonochromatic}, we consider non-monochromatic GW signals and explore the dynamics of mode functions for GW frequencies that grow linearly with time.
We derive the total energy deposited inside the cavity for transient signals, as an estimate for the experimental sensitivity of the apparatus.
We apply our results to PBH binary mergers and illustrate the experimental reach of a semi-realistic spherical cavity setup towards this scenario.
Finally, we briefly summarize and conclude in \cref{sec:conclusions}.

\section{Gravitational Wave Detection with Electromagnetic Cavities}
\label{sec:gwDetectionCavities}

The detection of GWs with electromagnetic cavities is based on the interaction between classical gravitational and electromagnetic fields, characterized by the Einstein-Maxwell action,
\begin{equation}
    S = \int \md^4 x \, \sqrt{-g} \left(-\frac{1}{4} g^{\mu \alpha} g^{\nu \beta} F_{\mu \nu} F_{\alpha \beta} \right) \, .
    \label{eq:actionEinsteinHilbert}
\end{equation}
It implies that, in the vicinity of an external magnetic field, a GW can be converted into electromagnetic radiation~\cite{Gertsenshtein:1962}, which, in turn, may be amplified inside an electromagnetic cavity to generate a detectable signal.
To characterize this signal, we assume that the GW is a weak gravitational perturbation of a flat Minkowski background,
\begin{equation}
    g_{\mu \nu} = \eta_{\mu \nu} + h_{\mu \nu} \, ,
    \label{eq:metricDecomposition}
\end{equation}
where $h_{\mu \nu}$ parametrizes the GW with $\absl{h_{\mu \nu}} \ll 1$.
Similarly, any tensor quantity can be systematically organized in powers of $h_{\mu \nu}$.
For instance, the electromagnetic field strength can be decomposed as $F_{\mu \nu} = \bar{F}_{\mu \nu} + \delta F_{\mu \nu}$, where the fluctuations $\delta F_{\mu \nu}$ are of order $\mathcal{O}(h)$ as compared to the background value $\bar{F}_{\mu \nu}$.

The perturbation scheme in \cref{eq:metricDecomposition} gives a prescription to solve Maxwell's equations in a curved spacetime at each order in the metric perturbation (see, e.g., \cite{Maggiore:2007ulw}).
While the homogeneous equations remain unaffected by the GW, the electromagnetic fluctuations satisfy the inhomogeneous equations
\begin{equation}
    \partial_{\nu} \delta F^{\mu \nu} = j_{\mathrm{eff}}^\mu \, ,
    \label{eq:maxwellLeadingOrder}
\end{equation}
where we have defined the effective current
\begin{equation}
    j_{\mathrm{eff}}^{\mu} = -\frac{1}{2} \left(\partial_{\lambda}h\right) \bar{F}^{\mu \lambda} + \partial_{\nu} \left( \tensor{h}{^\mu _\lambda} \bar{F}^{\lambda \nu} + \tensor{h}{^\nu _\lambda} \bar{F}^{\lambda \mu} \right) \, .
    \label{eq:effectiveCurrent}
\end{equation}
Here, $h = \tensor{h}{^\mu _\mu}$ denotes the trace of the metric perturbation.
\cref{eq:maxwellLeadingOrder} illustrates that an incoming GW sources fluctuations of the electromagnetic fields inside the cavity.

\subsection{The measured electromagnetic fields}

The electromagnetic fields governed by \cref{eq:maxwellLeadingOrder} are components of the field strength tensor $\delta F_{\mu \nu}$, which are clearly not invariant under coordinate transformations.
Therefore, they do not pose well-defined physical observables.
To define manifestly coordinate-independent expressions for the electromagnetic fields measured by an external observer, we closely follow~\cite{Ratzinger:2024spd} and associate an infinitesimal proper coordinate system to the observer's worldline, given by an orthonormal tetrad with coefficients $e_{\underline{\alpha}}^{\mu}$ in a given coordinate basis.
This relates the expressions for the observable electromagnetic fields to the frame-dependent motion of the detector.
For instance, to first order in the metric fluctuation, the observable coordinate-invariant electric field reads
\begin{equation}
    E_{\underline{a}}
    = \delta x^{\lambda} \left( \partial_{\lambda} \bar{F}_{\mu \nu} \right) \bar{e}^{\mu}_{\underline{a}} \bar{u}^{\nu}
    + \delta F_{\mu \nu} \bar{e}^{\mu}_{\underline{a}} \bar{u}^{\nu}
    + \bar{F}_{\mu \nu} \delta e^{\mu}_{\underline{a}} \bar{u}^{\nu}
    + \bar{F}_{\mu \nu} \bar{e}^{\mu}_{\underline{a}} \delta u^{\nu} \, .
    \label{eq:electricFieldObservable}
\end{equation}
Here, $\delta x^{\mu}$ denotes the perturbation of the observer's position,  and $u^{\mu}$ is the observer's four-velocity. 
In principle, all these contributions have to be taken into account to determine the observable electric field inside the cavity.
In practice, however, several terms do not contribute to the overall expression.
In our scenario, the background field is static and homogeneous, such that the first term vanishes, $\partial_{\lambda} \bar{F}_{\mu \nu} = 0$.
Furthermore, in our perturbation scheme, the background values of the tetrad are trivial, $\bar{e}_{\underline{\alpha}}^{\mu} = \delta^{\mu}_{\underline{\alpha}}$, and therefore $\bar{u}^{\nu} = \delta^{\nu}_{\underline{0}}$.
Since the background electric field vanishes, this term again does not contribute to the observable electric field.
Simplifying things even further, in a scenario where the GW frequency and the size of the detector are comparable, $\omega L \gtrsim 1$, the free-falling limit in TT gauge is a natural choice of frame and provides an excellent approximation to the exact result~\cite{Ratzinger:2024spd}.
In this limit the motion of the detector is suppressed by the sound speed $v_s$ inside the detector.
In other words, in a regime close to electromagnetic resonances of the cavity, we can safely neglect the motion of the detector, $\delta x^{\mathrm{TT}} \to 0$.
Therefore, the observable electric field finally reduces to the components of the electromagnetic field strength tensor in TT gauge,
\begin{equation}
    E_{i} \approx \delta F_{0 i}^{\mathrm{TT}} \, ,
\end{equation}
which differs from the exact result in \cref{eq:electricFieldObservable} by terms of the order $\mathcal{O}(v_s / (\omega L))$.
This is in contrast to earlier results, which instead have identified Fermi-normal (FN) coordinates as the most suitable frame (see, e.g., \cite{Berlin:2021txa, Fischer:2025mpz}).
We remark that this choice of frame, however, neglects the unsuppressed detector motion which has to be taken into account in FN coordinates.

\subsection{The coupling of the gravitational wave to the cavity modes}

The electric field that is excited inside the cavity is determined by Maxwell's equation given in \cref{eq:maxwellLeadingOrder}.
In the following, we perform our computations in TT gauge, dropping the TT superscript from now on.
The spatial components of the effective current sourced by an incoming GW are factorized into a temporal and a spatial contribution,
\begin{equation}
    \jeff ( \vec{x}, t) = \hat{\vec{j}} (\vec{x}) j_{\mathrm{eff}} (t) \, .
\end{equation}
In general, any electromagnetic cavity is sensitive to signals which can be resonantly amplified inside its volume.
Neglecting mechanical deformations, the excited electric field can be decomposed into the resonant modes $\vec{E}_n$,
\begin{equation}
	\vec{E}(\vec{x}, t) = \sum_{n=0}^{\infty} \vec{E}_n(\vec{x}) e_n(t) \, ,
	\label{eq:electricFieldModeDecomposition}
\end{equation}
where the time-dependence of the field is separated into the dimensionless mode functions $e_n(t)$ of mode number $n$.
The modes are characterized by the resonant frequencies $\omega_n$ and form an orthogonal basis of the spectrum (see, e.g., \cite{hill2009electromagnetic}),
\begin{align}
	\nabla^2 \vec{E}_n &= -\omega_n^2 \vec{E}_n \, , \label{eq:emModeFunctions1} \\
	\int_{\vcav} \md^3 x \, \vec{E}_m^{\ast} \cdot \vec{E}_n &= \delta_{mn} \int_{\vcav} \md^3 x \, \abs{\vec{E}_n}^2 \, , \label{eq:emModeFunctions2}
\end{align}
where $\vcav$ is the volume of the cavity.
Since the homogeneous equations are not affected by the incoming GW in our perturbation scheme, the electric field satisfies an electromagnetic wave equation with the effective current as a source term, such that the mode functions obey~\cite{Berlin:2021txa}
\begin{equation}
	\left( \partial_t^2 + \frac{\omega_n}{Q_n} \partial_t + \omega_n^2 \right) e_n (t) = \xi \partial_t j_{\mathrm{eff}} \, .
	\label{eq:eomModeFunctions}
\end{equation}
Here, we have introduced the (mode-dependent) quality factor $Q_n$ of a cavity, which parametrizes energy losses of the field inside the cavity.
Furthermore, we have defined the coupling constant
\begin{equation}
	\xi = -\frac{\int_{\vcav} \md^3 x \, \vec{E}_n^{\ast} \cdot \hat{\vec{j}}}{\int_{\vcav} \md^3 x \, \abs{\vec{E}_n}^2} \, .
\end{equation}
Clearly, \cref{eq:eomModeFunctions} describes a harmonic oscillator with frequency $\omega_n$ and damping $1 / Q_n$, which is driven by the oscillations of the effective current induced by the incoming GW.

While the spatial modes $\vec{E}_n$ are determined by the cavity geometry (see, e.g., \cite{hill2009electromagnetic}), the mode functions $e_n$ are governed by the time-dependence of the effective current.
Assuming that the incoming GW is monochromatic and of frequency $\omega_g \simeq \omega_n$, inducing an effective current with $j_{\mathrm{eff}}(t) = \exp (i \omega_g t)$, the mode functions can be calculated explicitly,
\begin{equation}
    e_n(t) \simeq \xi \frac{Q_n}{\omega_g} \me^{i \omega_g t} \, .
    \label{eq:modeFunctionMonochromatic}
\end{equation}
Here, $\xi$ captures the coupling of the incoming GW to the resonant modes of the cavity.
The measured signal will be proportional to the energy density, $\sim \absl{\vec{E}}^2$, inside the cavity volume, and hence it will behave as $\propto \xi^2$.
To estimate this signal sensitivity in terms of a \emph{dimensionless} coupling constant, we follow~\cite{Berlin:2021txa} and decompose the spatial contribution to the effective current into dimensionless components $\hat{\vec{j}}_{+} (\vec{x})$ and $\hat{\vec{j}} (\vec{x})_{\times}$, such that
\begin{equation}
    \hat{\vec{j}} (\vec{x}) = \omega_g B_0 \left( h_{+} \hat{\vec{j}}_{+} (\vec{x}) + h_{\times} \hat{\vec{j}}_{\times} (\vec{x}) \right) \, .
    \label{eq:effectiveCurrentDimensionless}
\end{equation}
Here, $B_0$ is the homogeneous background magnetic field, and $h_{+}$ and $h_{\times}$ denote the different polarization coefficients of the incoming GW.
We remark that, as we will see later, this result naturally holds in TT gauge.
In contrast, this decomposition comes with an additional factor $\omega_g \vcav^{1/3}$ if FN coordinates are used~\cite{Berlin:2021txa}, because both frames scale differently with $\omega L$ (see, e.g.,~\cite{Ratzinger:2024spd}).
Using \cref{eq:effectiveCurrentDimensionless}, we can define a dimensionless coupling coefficient of the incoming GW to the $n$-th mode of the electromagnetic cavity~\cite{Berlin:2021txa},
\begin{equation}
    \eta_n^{+, \times} = \frac{\abs{\int_{\vcav} \md^3 x \, \vec{E}_n^{\ast} \cdot \hat{\vec{j}}_{+,\times}}}{\sqrt{\vcav \int_{\vcav} \md^3 x \, \abs{\vec{E}_n}^2}} \, .
    \label{eq:couplingCoefficient}
\end{equation}
Here, we have focused on a single mode with index $n$ that contributes to the total observable electric field.
Some modes may be degenerate however, so that they can in principle be excited simultaneously by the same incoming GW.
These need to be carefully taken into account when determining the observable electric field.

\subsection{The power output at the antenna}

Let us now turn our attention to a characterization of the experimental sensitivity of an electromagnetic cavity to a GW signal.
Once the incoming GW has deposited energy inside the cavity, the excited electric and magnetic fields oscillate at their resonant frequencies according to the boundary conditions of the detector material.
These oscillations can be measured by an antenna placed inside the cavity volume.
Clearly, this requires a coupling of the antenna to the oscillating electromagnetic fields, such that some amount of energy dissipates at the antenna.
In an experimental setup, this energy dissipation is indeed a measurement observable in terms of a power output at the antenna, which is schematically given by
\begin{equation}
    P_{\mathrm{out}} = -\frac{\md}{\md t} W \, .
\end{equation}
Here, $W$ denotes the (time-averaged) total energy stored inside the cavity.
Following Poynting's theorem, it is given by the integral the electromagnetic fields over the cavity volume,
\begin{equation}
    W = \frac{1}{2} \int_{\vcav} \md^3 x \, \abs{\vec{E}(\vec{x}, t)}^2 \, .
    \label{eq:energyElectricField}
\end{equation}
This means, once a certain set of electromagnetic cavity modes is occupied, an observer would measure the total energy loss of the cavity over time.
On the other hand, energy may not only be dissipated at the antenna, but there are also energy losses due to the absorption of photons a the cavity walls.
These losses are typically characterized by the quality factor of a cavity,
\begin{equation}
    Q = \left(\frac{\delta W}{\delta n_c}\right)^{-1} W \, ,
    \label{eq:qualityFactorDefinition}
\end{equation}
which relates the total energy stored in the cavity to the amount of energy that is dissipated per oscillation cycle, $\delta W / \delta n_c$.
The latter can be expressed as $\delta W / \delta n_c \sim \delta W / (\omega_n \delta t)$, i.e.~it is itself proportional to the change in energy over time.
Therefore, \cref{eq:qualityFactorDefinition} poses a first-order differential equation describing the exponential decay of a cavity's energy due to dissipation effects.
This, of course, assumes that $Q$ is constant and \emph{defines} the decay constant of this energy dissipation, which may be determined by averaging over many oscillation cycles.
Therefore, \cref{eq:qualityFactorDefinition} implies that the total power loss of a given cavity mode reads
\begin{equation}
    P = \frac{\omega_n}{Q_n} W \, .
    \label{eq:powerOutputTimeAverage}
\end{equation}
We stress that $P$ parametrizes the total loss of energy of the cavity, including the losses due to the cavity walls, $P_{\mathrm{cav}}$ as well as the antenna readout, $P_{\mathrm{out}}$, such that $P = P_{\mathrm{cav}} + P_{\mathrm{out}}$.
Experimentally, both sources of power output are typically tuned by a parameter $\beta$ to be roughly of the same order of magnitude, $P_{\mathrm{out}} = \beta P_{\mathrm{cav}}$.
In practice, one often has $\beta \simeq 2$.
The power loss measured at the antenna therefore reads
\begin{equation}
    P_{\mathrm{out}} = \frac{\omega_n}{\left(1 + \frac{1}{\beta}\right) Q_n} W \, .
\end{equation}
Finally, using the dimensionless coupling coefficients defined in \cref{eq:couplingCoefficient}, the power output of a monochromatic GW signal can be written as
\begin{equation}
    P_{\mathrm{out}} = \frac{Q}{2 \left(1 + \frac{1}{\beta}\right)} \omega_g \vcav \left( \eta_n^{+,\times} B_0 h_{+,\times} \right)^2 \, .
    \label{eq:powerOutputCoupling}
\end{equation}
Here, for simplicity, we have assumed that only one polarization is contributing to the effective current.

Before we close this discussion, let us briefly comment on how the coupling coefficients associated to each mode are related to each other.
Naively, if the cavity is invariant under a set of symmetry transformations (such as rotations, for example), we expect the spectrum to be degenerate.
That is, an incoming GW of a given frequency $\omega_g \simeq \omega_n$ may excite various different modes simultaneously.
So far, this is not captured by the coupling coefficients $\eta^{+,\times}$ defined in \cref{eq:couplingCoefficient}.
Instead, as has been noted in~\cite{Berlin:2021txa}, to account for a degenerate spectrum, we have to include all mode functions of a given frequency.
To illustrate this, suppose that two modes $n$ and $m$ are degenerate, $\omega_n = \omega_m$, and are excited by an incoming monochromatic GW of the same frequency.
According to \cref{eq:electricFieldModeDecomposition}, this will lead to an electric field inside the cavity, which is of the form
\begin{equation}
    \vec{E}(\vec{x},t) \simeq \frac{Q}{\omega_g} \left( \xi_n \vec{E}_n (\vec{x}) + \xi_m \vec{E}_m (\vec{x}) \right) \me^{i \omega_g t} \, .
\end{equation}
The total power output of the cavity is proportional to the total energy stored in the field, $P_{\mathrm{out}} \propto \int_{\vcav} \absl{\vec{E}}^2$, which clearly contains interference terms between the $n$-th and $m$-th cavity mode.
However, by construction of \cref{eq:emModeFunctions2}, these are orthogonal.
Therefore, the overall coupling coefficient introduced to characterize the power output is
\begin{equation}
    \eta_{\mathrm{tot}}^{+, \times} = \sqrt{\left(\eta_n^{+,\times}\right)^2 + \left(\eta_m^{+,\times}\right)^2} \, .
    \label{eq:couplingCoefficientDegenerate}
\end{equation}
This result can be straightforwardly generalized to a scenario involving an arbitrarily degenerate spectrum.
Let us now use \cref{eq:couplingCoefficientDegenerate} to characterize the electromagnetic signal strength of an incoming monochromatic GW to different cavity geometries.

\section{Characterizing the Electromagnetic Signal}
\label{sec:couplingCoefficients}

Similar to axion searches, an electromagnetic cavity is typically prepared in the vicinity of a background magnetic field to convert a gravitational into an electromagnetic wave inside the cavity volume.
In the following, we choose this external magnetic field to be homogeneous and pointing along the $z$-axis,
\begin{equation}
    \bar{\vec{B}} = B_0 \vec{e}_z \, .
    \label{eq:magneticFieldBackground}
\end{equation}
Further suppose that the incoming GW is monochromatic with frequency $\omega$ and has an incidence angle $\alpha$ with respect to the magnetic field.
For concreteness, we choose a coordinate system where the GW is propagating in the $yz$-plane with wave vector $\vec{k} = \omega \left(0, \sin \alpha, \cos \alpha \right)$.
As we have briefly discussed in \cref{sec:gwDetectionCavities}, in practice, the free-falling limit provides an excellent approximation to characterize the resonant electromagnetic excitations of the cavity due to the GW~\cite{Ratzinger:2024spd}.
Hence, we parametrize the GW in TT gauge (see, e.g., \cite{Domcke:2022rgu}),
\begin{equation}
    h_{ij}^{\mathrm{TT}} = \frac{1}{\sqrt{2}}
        \left( h_{+} \mathbb{M} + h_{\times} \mathbb{N} \right)
        \me^{i \omega \left( -t + y \sin \alpha + z \cos \alpha \right)} \, ,
    \label{eq:hTTGauge}
\end{equation}
where $h_{+}$ and $h_{\times}$ are the different polarization amplitudes, and we have defined
\begin{align}
    \mathbb{M} = \begin{pmatrix}
                -1 &             0 &                        0 \\
                 0 & \cos^2 \alpha & -\cos \alpha \sin \alpha \\
                 0 & -\cos \alpha \sin \alpha & \sin^2 \alpha
            \end{pmatrix} \, , \enspace
    \mathbb{N} = \begin{pmatrix}
                           0 & -\cos \alpha & \sin \alpha \\
                -\cos \alpha &            0 &           0 \\
                 \sin \alpha &            0 &           0
            \end{pmatrix} \, .
\end{align}
The electromagnetic response of the cavity to the incoming GW is governed by \cref{eq:maxwellLeadingOrder}, and is driven by the effective current induced by $h_{ij}$.
Plugging both the external magnetic field and the GW parametrization into \cref{eq:effectiveCurrent}, the spatial components of the effective current read\footnote{By comparing to axion electrodynamics, a convenient way to determine the effective current in a given coordinate frame is provided in~\cite{Domcke:2022rgu}.}
\begin{equation}
    \jeff (\vec{x}, t) =
        -\frac{i}{\sqrt{2}} \omega_g B_0
        \begin{pmatrix}
            h_{+} \sin \alpha \\
            h_{\times} \sin \alpha \cos \alpha \\
            -h_{\times} \sin^2 \alpha
        \end{pmatrix}
        \me^{i \omega_g \left( -t + y \sin \alpha + z \cos \alpha \right)} \, .
\end{equation}
Clearly, $\jeff$ will act as a harmonically oscillating driving force in the equations of motion of the induced electromagnetic fields inside the cavity.

The experimental sensitivity to a monochromatic GW is characterized by the signal power output of the cavity given in \cref{eq:powerOutputCoupling}.
Clearly, beyond the background magnetic field $B_0$ as well as the cavity's volume $\vcav$ and quality factor $Q$, this power output is governed by the GW's effective coupling to the various resonant excitations, $\eta_n^{+,\times}$ introduced in \cref{eq:couplingCoefficient}.
In the following, we illustrate this effective coupling coefficient for selected example configurations involving cylindrical as well as spherical cavities.

\subsection{Cylindrical cavities}
\label{sec:cylindricalCavity}

Cylindrical cavities are an established technology used for axion searches, making them a natural candidate to study their experimental sensitivity to high-frequency GWs.
To do this, let us first follow the example discussed in~\cite{Berlin:2021txa} and consider a cylindrical cavity of radius $R$ and length $L$.
For simplicity, we choose the cylinder's rotation axis to be along the $z$-axis, thereby aligning with the external magnetic field.

In general, the resonant modes of a cavity are typically given in terms of transverse electric (TE) and transverse magnetic (TM) fields.
For a cylindrical cavity, these are characterized by three quantum numbers~\cite{hill2009electromagnetic}, schematically denoted by $\vec{E}_{npq}$.
We present the expressions for the resonant electric field configurations in \cref{app:modesCylindricalCavity}.
While the resonance frequencies depend on all three quantum numbers $n$, $p$, and $q$, the mode configurations feature both a contribution that is even or odd in $n$, i.e.~they are doubly-degenerate~\cite{hill2009electromagnetic}.
According to \cref{eq:couplingCoefficientDegenerate}, the total coupling coefficient of the GW to $\vec{E}_{npq}$ has to account for this degeneracy,
\begin{equation}
    \eta_{npq}^{+,\times} = \sqrt{\left( \eta_{npq, \mathrm{even}}^{+,\times} \right)^2
                                + \left( \eta_{npq, \mathrm{odd}}^{+,\times} \right)^2} \, .
\end{equation}
Here, the superscript denotes the GW polarization.

\begin{figure}[t]
	\centering
    \includegraphics{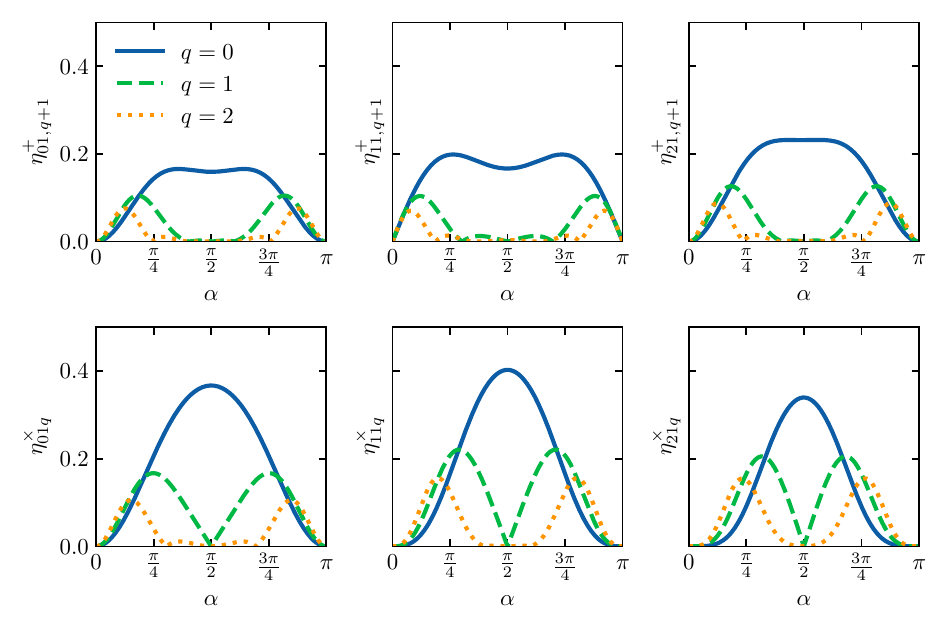}
	\caption{Coupling coefficients $\eta^{+,\times}_{npq}$ of the TE (top) and TM modes (bottom) of a cylindrical cavity as a function of the incidence angle $\alpha$ of the incoming GW, parametrized in TT gauge. The different colors illustrate the longitudinal quantum numbers $q=0$ (solid blue), $q=1$ (dashed green), and $q=2$ (dotted orange), while the principal quantum number is fixed, $p=1$. In both rows, the azimuthal quantum number $n$ is increased to the right, with $n=0$ (left), $n=1$ (center), and $n=2$ (right). Note that the cylinder dimensions are chosen such that $R=L$, and that different polarizations are shown for the top and bottom panels.}
	\label{fig:etasCylindrical}
\end{figure}

The combinations of quantum numbers associated to the resonant modes with the largest $\eta_{npq}$ are illustrated in \cref{fig:etasCylindrical}.
Here, we show the coupling coefficients for various low-lying modes as a function of the GW incidence angle $\alpha$ with respect to the background magnetic field.
We find that, for both TE and TM modes, the longitudinal quantum number $q$ has a strong impact on the overall magnitude of the coupling.
For instance, the number of roots of $\eta^{+,\times}_{npq}$ increases while their overall magnitude rapidly decreases for larger $q$, suggesting that the lowest-lying cavity modes exhibit the strongest coupling to GWs.
The maximum coupling strength at $q=0$ is typically achieved for an incoming GW that is propagating perpendicular to the background magnetic field, $\alpha = \pi/2$.
On the other hand, we also note that $\eta^{+,\times}_{npq}$ vanishes at $\alpha = 0$ and $\alpha = \pi$, such that one would not expect a signal induced by a GW propagating parallel or antiparallel to the magnetic field.
We remark that this is somewhat in contrast to coupling coefficients obtained in FN coordinates~\cite{Berlin:2021txa} (see also~\cite{Fischer:2025mpz}), for which we provide a brief comparison in \cref{app:fnCoordinates}.

Finally, we note that \cref{fig:etasCylindrical} illustrates the coupling of the TE and TM modes to certain polarization amplitudes of the GW only; the combinations of polarizations that are not shown here are considerably smaller.
Overall, the TM modes exhibit the largest coupling to the cross-polarization of an incoming monochromatic GW.
In particular, the resonant TM$_{010}$ mode, which is often used for axion searches in practice, features a comparably large coupling coefficient.
Given currently available experimental technology and expertise it may therefore be the most suitable mode to consider for a GW search using cylindrical cavities.

\subsection{Spherical cavities}
\label{sec:sphericalCavity}

The only preferred direction in a setup involving a spherical cavity is given by the external magnetic field.
Given the high amount of symmetry, we naturally expect a degenerate spectrum of modes.

\begin{figure}[t]
	\centering
    \includegraphics{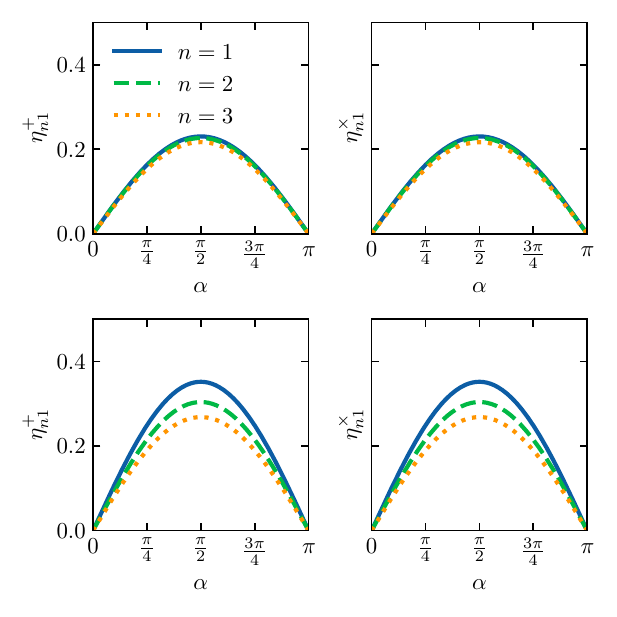}
	\caption{Coupling coefficients $\eta^{+,\times}_{np}$ of the TE (top) and TM modes (bottom) of a spherical cavity as a function of the incidence angle $\alpha$ of the incoming GW, parametrized in the TT frame. The different colors illustrate the azimuthal quantum numbers $n=1$ (solid blue), $n=2$ (dashed green), and $n=3$ (dotted orange), while the principal quantum number is fixed, $p=1$. The modes are $(2n+1)$-fold degenerate with respect to the azimuthal quantum number, which is hence marginalized in this example (see main text).}
	\label{fig:etasSpherical}
\end{figure}

The resonant modes of a spherical cavity are also characterized by three quantum numbers, $\vec{E}_{mnp}$~\cite{hill2009electromagnetic}.
Analogous to the quantum mechanical wave function of the hydrogen atom, $m$ corresponds to the magnetic, $n$ denotes the azimuthal, and $p$ corresponds to the principal quantum number.
The associated field configurations are given in \cref{app:modesSphericalCavity}.
The resonant mode frequencies, however, only depend on $n$ and $p$.
The magnetic quantum number takes values $m \in \{-n, \ldots, n\}$, such that each mode with azimuthal quantum number $n$ is $(2n+1)$-fold degenerate.
Consequently, the total coupling coefficient reads
\begin{equation}
    \eta_{np}^{+,\times} = \sqrt{\sum_{m=-n}^n \left( \eta_{mnp}^{+,\times} \right)^2} \, ,
\end{equation}
where the superscript denotes the GW polarization.

The coupling coefficients of a spherical cavity are shown in \cref{fig:etasSpherical}, where we illustrate combinations of quantum numbers associated to the TE and TM modes featuring the largest $\eta_{np}$ as a function of the GW incidence angle $\alpha$.
We find that, generically, modes with $p=1$ exhibit the strongest coupling, which decreases rapidly for larger $p$.
The lowest-lying modes in the azimuthal quantum number $n$ are the dominant ones.
Similar to the cylindrical cavity, the largest coupling is obtained for an incoming GW that traverses the cavity perpendicular to the magnetic field, at $\alpha = \pi / 2$.
On the other hand, the coupling coefficient vanishes at $\alpha = 0$ and $\alpha = \pi$, leading to a vanishing signal for a GW that propagates parallel or antiparallel to the external magnetic field.
We again remark that these results are presented in TT gauge which we believe is the most suitable choice of coordinates~\cite{Ratzinger:2024spd}.
In \cref{app:fnCoordinates} we provide a brief comparison to FN coordinates.

In summary, we find that the coupling of the incoming GW to the dominant cavity modes is sizable, and comparable to the case of a cylindrical cavity illustrated in \cref{fig:etasCylindrical}.
Overall, for a maximum signal strength the incoming GW would have to propagate perpendicular to the external magnetic field.
As a blessing in disguise, we can turn this argument around to obtain an optimal sky coverage, for instance by preparing more than one cavity each with an external magnetic field pointing into a different direction.
At the same time, this would enable performing coincidence measurements of potential GW signals, somewhat similar to the idea behind the ``GravNet" proposal~\cite{Schmieden:2023fzn}.

\subsection {Sensitivity prospects}

\begin{figure}[t]
	\centering
	\includegraphics{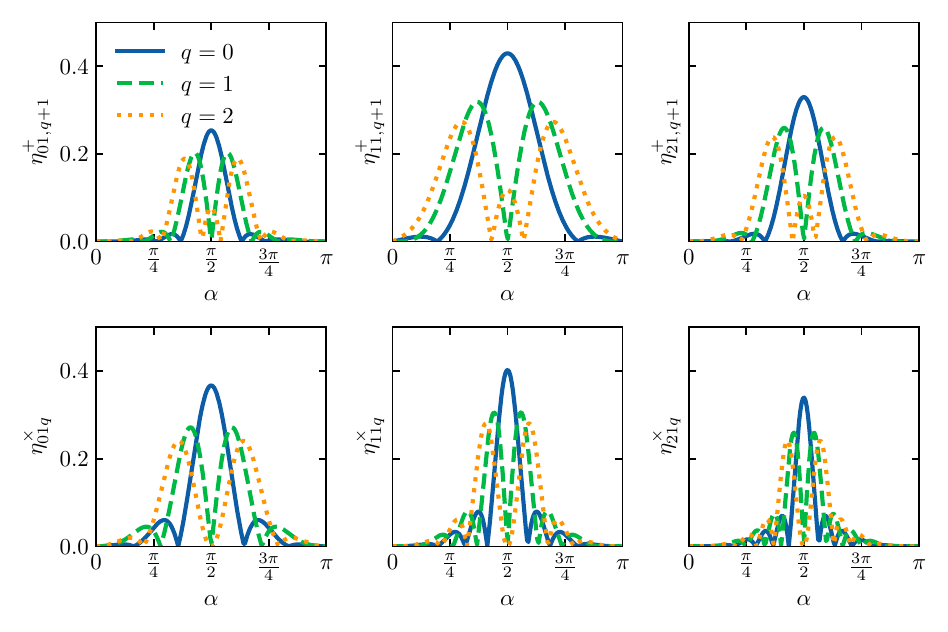}
	\caption{Similar to \cref{fig:etasCylindrical}, but with dimensions $L / R = 6$.}
	\label{fig:etasCylindricalSupax}
\end{figure}

A typical cavity experiment to search for GWs in the GHz range will utilize a cryogenic setup involving an available cylindrical volume with radius $R=2.5$~cm and length $L=15$~cm, prepared in the vicinity of an external magnetic field with $B=12$~T.
Changing the ratio between the cavity's radius and length drastically alters its angular sensitivity.
This is illustrated in \cref{fig:etasCylindricalSupax}, where we show the dimensionless coupling coefficients for a cylindrical cavity with $L/R = 6$ (in contrast to \cref{fig:etasCylindrical} which uses $L/R = 1$).
To estimate the expected sensitivity of this setup, we first determine the power induced by a monochromatic GW using \cref{eq:powerOutputCoupling}.
The corresponding signal-to-noise ratio is given by
\begin{equation}
    \mathrm{SNR} \simeq \frac{P_{\mathrm{out}}}{T_{\mathrm{sys}}} \sqrt{\frac{2\pi t_{\mathrm{int}}}{\Delta \omega}} \, ,
    \label{eq:snr}
\end{equation}
where $T_{\mathrm{sys}}$ is the effective thermal noise temperature, $t_{\mathrm{int}}$ denotes the measurement integration time, and $\Delta \omega$ is the frequency bandwidth of the measurement.
For a detectable GW strain we require a signal-to-noise ratio of $\mathrm{SNR} \gtrsim 1$ and assume a thermal noise temperature of $T_{\mathrm{sys}} = 200$~mK, which is typically limited by the system noise temperature of the amplifier.
As illustrated in \cref{tab:cavityParameters}, for integration times of $t_{\mathrm{int}} = 1$~min, GW strain sensitivities of the order $\mathcal{O}(10^{-20})$ can be reached with this setup, assuming an effective coupling coefficient of the order $\eta \simeq 0.75 \, \eta_{\mathrm{max}}$.
For comparison, we also illustrate an analogous setup where three spherical cavities of the same radius, or six cylindrical cavities are stacked on top of each other to obtain the same available detection volume.
The corresponding coupling coefficients and angular sensitivity are shown in \cref{fig:etasCylindrical,fig:etasSpherical}.
For integration times of the order of minutes, we find that all three setups are equally sensitive.
This changes once we account for the angular sensitivity of the different cavities, illustrated by the opening angle $\angle (\eta_{0.5})$ in \cref{tab:cavityParameters}, as we will see momentarily.

\begin{table}[t]
    \centering
    \begin{tabular}{cccc}
    \toprule
         &  Spherical & \multicolumn{2}{c}{Cylindrical}\\
    \midrule
        $R$ [cm] & 2.5 & 2.5 & 2.5 \\
        $L$ [cm] & --- & 2.5 & 15.0 \\
        $\vcav$ [cm$^3$]& $3 \times 65.4$ & $6 \times 49.1$ & 295.0 \\
        $\omega$ [GHz] & 32.9 & 28.8 & 28.8 \\
        $\eta_{\mathrm{max}}$ & 0.35 & 0.37 & 0.37 \\
        $\angle (\eta_{0.5})$ & $120^{\circ}$ & $96^{\circ}$ & $30^{\circ}$ \\
    \midrule
        \multicolumn{4}{c}{$t_{\mathrm{int}} = 1$~min} \\
        \midrule
        $h_0$ & $1.0 \times 10^{-20}$ & $8.1 \times 10^{-21}$ & $8.1 \times 10^{-21}$ \\
        $r$ [pc] & 0.09 & 0.13 & 0.13 \\
    \midrule
    \multicolumn{4}{c}{$t_{\mathrm{int}} = \angle (\eta_{0.5}) / 360^{\circ}$~day} \\
    \midrule 
        $h_0$ & $2.2 \times 10^{-21}$ & $1.83 \times 10^{-21}$ & $2.5 \times 10^{-21}$ \\
        $r$ [pc] & 0.41 & 0.55 & 0.41 \\
    \bottomrule
    \end{tabular}
    \caption{Overview of typical parameters for a spherical and cylindrical cavity setup. The dominant modes $\omega_{11}^{\mathrm{TM}}$ (spherical) and $\omega_{010}^{\mathrm{TM}}$ (cylindrical) are illustrated, for which the coupling constant $\eta$ is maximal. Furthermore, the opening angle for which the coupling is larger than half of the maximum $\eta_{\mathrm{max}}$ is shown, $\angle (\eta_{0.5})$. In addition, the GW strain $h_0$ to obtain a signal-to-noise ratio of $\mathrm{SNR}=1$ is illustrated, assuming an external magnetic field of $B=12$~T, a 200~mK thermal background and a quality factor of $Q = 6 \times 10^4$. For monochromatic GWs sourced by superradiance this can in turn be translated into a distance $r$ to the source.}
    \label{tab:cavityParameters}
\end{table}

A common benchmark for coherent monochromatic GW sources in the GHz range are boson clouds created by superradiant instabilities around spinning (primordial) black holes~\cite{Brito:2015oca, Aggarwal:2025noe}.
The corresponding GW frequency and strain are given by 
\begin{align}
    \omega_g &\approx 15~{\rm GHz} \left(\frac{10^{-5} M_\odot}{m_{\rm PBH}}\right) \left( \frac{G m_{\rm PBH}m_\phi}{0.1} \right) \, , \\
    h_0 &\approx 5 \times 10^{-30}\left(\frac{m_{\rm PBH}}{10^{-5} M_\odot}\right)\left( \frac{G m_{\rm PBH}m_\phi}{0.1} \right)^7 \left( \frac{{\rm kpc}}{r}\right) \, ,
\end{align}
where $m_{\rm PBH}$ denotes the PBH mass, $G$ the gravitational constant, $m_\phi$ the mass of the light bosonic field, and $r$ the distance of the source from the detector.
Superradiance requires that $2G m_{\mathrm{PBH}} m_\phi \sim 1$, such that, effectively, the emitted GW frequency is set by the PBH mass alone.
For typical measurement integration times of a minute the example setup given in \cref{tab:cavityParameters} can probe distances up to $r=0.14$~pc, corresponding to PBH masses around $m_{\mathrm{PBH}} \approx 2.5 \times 10^{-5} M_{\odot}$.
Clearly, the strain sensitivity can be improved by increasing the integration time $t_{\mathrm{int}}$, albeit it only grows as $\sim t_{\mathrm{int}}^{1/4}$.
We illustrate a second benchmark point in \cref{tab:cavityParameters} where we, as a naive estimate, increase the integration time from one minute to the time span during which the GW reaches the detector within its opening angle, such that the coupling coefficient exceeds half its maximum over the course of a day.
This leads to an increase in the sensitivity by roughly a factor of four, while the stacked setups now perform better for the resonant modes considered here.

\begin{figure}[t]
	\centering
	\includegraphics{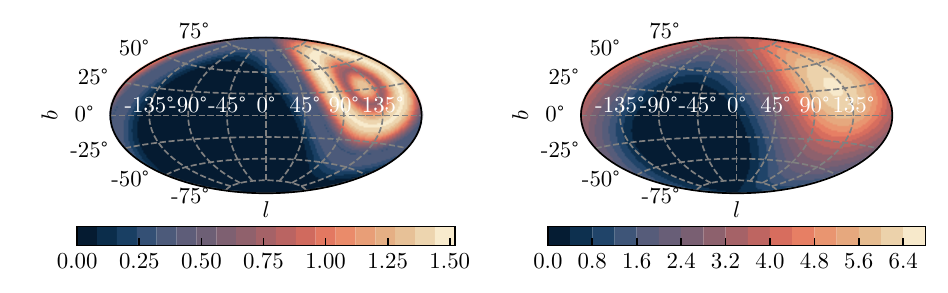}
	\caption{Integrated coupling coefficient, $\int \md t \, \eta$, of a monochromatic GW source corresponding to a low-lying TM mode of a cylindrical (left) and a spherical (right) cavity, as a function of galactic longitude $l$ and latitude $b$. Both experiments are assumed to be set up in Mainz, Germany, with an external magnetic field normal to the Earth's surface and a measurement integration period of 24 hours. For the geometry of the cylindrical cavity we assume $L/R = 6$. Note that the colored contours are shown in arbitrary units.}
	\label{fig:skyMap}
\end{figure}

More precisely, within one day of measurement time, the rotation of the Earth will allow to scan a band across the sky, which depends on the latitude of the detector and the directionality of the given cavity mode.
In practice, this enhances the sensitivity to monochromatic sources located in certain patches of the sky.
An example of this is shown in \cref{fig:skyMap}, where we illustrate the dimensionless coupling coefficient integrated over the course of 24 hours for GWs emitted by a source located at a given galactic longitude and latitude.
Here, the coupling coefficients correspond to TM modes $\omega_{010}^{\mathrm{TM}}$ for the cylindrical setup with $L/R = 6$, and $\omega_{11}^{\mathrm{TM}}$ for the spherical cavity.
As anticipated earlier, we find that spherical cavity modes have an advantage over cylindrical ones, as they allow to cover a broader patch of the sky.

Let us close this discussion with a few words of caution. As discussed above, the strain sensitivity growth with the integration time as $\sim t_{\mathrm{int}}^{1/4}$, so for an order-of-magnitude improvement one would need to increase the integration time by a factor of $10^4$.
At the same time, for high quality factors the frequency bandwidth is narrow, making the exploration of sizable fractions of parameter space expensive.
Clearly, this is not feasible.
Furthermore, this analysis relies on the incoming GWs being monochromatic.
This is indeed a somewhat optimistic scenario as we will elucidate in the following section.

\section{Beyond Monochromatic Gravitational Waves}
\label{sec:beyondMonochromatic}

We now go beyond the case of monochromatic GWs and consider a scenario where the incoming signal changes its frequency with time and scans through the different resonance bands of the cavity.
Let us first discuss the general dynamics of the electromagnetic excitations for non-monochromatic signals, before exploring their consequences for the case of PBH binary systems.

\subsection{Mode function dynamics}

Non-monochromatic GW signals induce an effective current with a time-dependent frequency, for example
\begin{equation}
	\jeff (t) \propto \me^{i \omega_g(t) t} \, .
	\label{eq:effectiveCurrentTimeDependent}
\end{equation}
Following \cref{sec:gwDetectionCavities}, let us illustrate an example of the electromagnetic field dynamics inside the cavity sourced by an effective current of the form given in \cref{eq:effectiveCurrentTimeDependent}.
For concreteness, we assume an incoming GW with a frequency that grows linearly with time,
\begin{equation}
	\omega_g(t) = \chi \left(t - t_n\right) + \omega_n \, .
    \label{eq:omega_of_t_1}
\end{equation}
In this parametrization, the GW changes its frequency at a rate $\chi$ before it hits the cavity's resonant frequency, $\omega_n$, at a time $t_n$.
The cavity's response to the incoming GW is governed by the equations of motion for the mode functions $e_n(t)$ in \cref{eq:eomModeFunctions}.
Considering a dimensionless time parameter, $\tau = \omega_n t$, these can be written as
\begin{equation}
	\left( \partial_{\tau}^2 + \frac{1}{Q} \partial_{\tau} + 1 \right) \epsilon_n (\tau) = \partial_{\tau} j \, ,
    \label{eq:cavity_oscillator}
\end{equation}
Here, we have defined $\epsilon_n = \omega_n e_n / \xi$.
Clearly, the dynamics are described by a driven, damped harmonic oscillator.

\begin{figure}[t]
	\centering
	\includegraphics{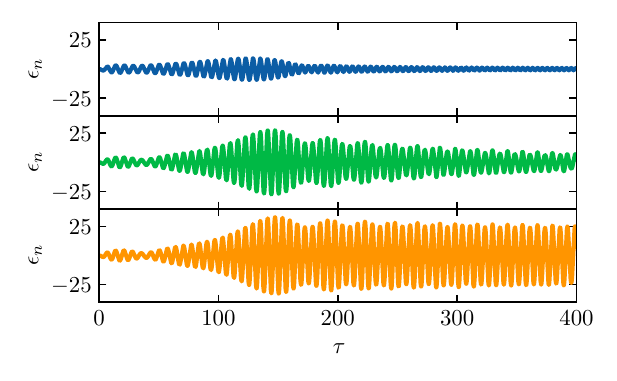}
	\caption{Rescaled mode functions $\epsilon_n$ as a function of dimensionless time $\tau$, for different quality factors, $Q_n=10$ (top), $Q_n=100$ (center), and $Q_n=1000$ (bottom). The scanning rate is chosen as $\chi / \omega_n^2 = 10^{-3}$ and the cavity's resonant frequency is reached at $\tau_n = 200$.}
	\label{fig:modeFunctionsDynamics}
\end{figure}

We illustrate an example of these dynamics in \cref{fig:modeFunctionsDynamics}, for different values of the quality factor $Q$.
As the effective current acts as an oscillating driving force that slowly approaches the cavity's resonant frequency, the oscillations of the mode functions are amplified steadily.
This happens as long as the driving force and the mode functions oscillate in phase.
In the example above, the resonant amplification sets in at $\tau \simeq \tau_n / 2$.
As the quality factor $Q$ is increased, the maximum amplitude of the oscillations at resonant frequency $\omega_n$ grows as $A_{\mathrm{max}} \sim Q / (\sqrt{\chi} \omega_n)$.
However, while the incoming GW signal traverses the cavity's resonance band, the relative phase between the cavity modes and the driving force grows, such that eventually both oscillations are no longer in phase and cease to be coherent.
The time scale after which decoherence occurs can be estimated as
\begin{equation}
    t_{\mathrm{decoherence}} \simeq \sqrt{\frac{\pi}{\chi}} \, .
\end{equation}
Therefore, there is a maximum number of oscillations $\nmax$ during which the mode function amplitude grows, and energy can accumulate inside the cavity,
\begin{equation}
    \nmax \simeq \sqrt{\frac{\omega_n^2}{4\pi \chi}} \, .
\end{equation}
This means that there is a critical value of the quality factor beyond which no further amplification occurs.
Instead, the system saturates because it is dominated by the maximum number of oscillations before decoherence sets in.
Finally, after the resonant frequency band is traversed the oscillations decay exponentially, $A \sim \exp(-\tau / Q)$.
We provide a detailed derivation and discussion of these dynamics in \cref{app:DrivenHarmonicOscillator}.

\begin{figure}[t]
	\centering
    \includegraphics{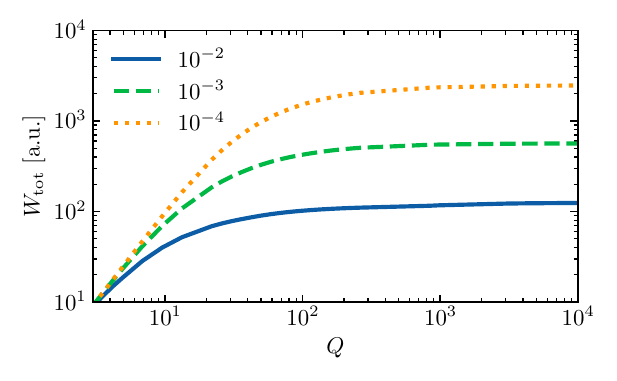}
	\caption{Maximum stored energy $\wtot$ for a loaded cavity as a function of the cavity's quality factor $Q$. The energy is shown in arbitrary units, while the different colors represent different dimensionless scanning rates $\chi / \omega_n^2$.}
	\label{fig:powerOutput}
\end{figure}

In other words, increasing the quality factor of the cavity indefinitely will not increase the cavity's response to the incoming non-monochromatic GW beyond a certain maximum value.
Let us estimate this effect by considering a simple setup where an incoming GW deposits a certain amount of energy inside the cavity.
At the same time, this amount of energy precisely corresponds to the maximum energy that an antenna can read out to detect the GW signal.
In general, following \cref{eq:energyElectricField}, the total energy should be proportional to the maximum amplitude of the oscillations, $\wtot \propto A_{\mathrm{max}}^2$.
According to our earlier discussion, for cavities with a large quality factor, $Q \gg \nmax$, the maximum amplitude of the oscillations is reached after $\nmax$ oscillations (see also \cref{app:DrivenHarmonicOscillator}).
In other words, in this regime, the total energy that can at most be deposited inside the cavity volume is limited by $\nmax$, otherwise is limited by the quality factor $Q$.
We find
\begin{equation}
    \wtot = \frac{1}{2} \int_{\vcav} \md^3 x \, \abs{\vec{E(\vec{x})}}^2 \simeq \frac{1}{2} \vcav B_0^2 h^2 \eta_n^2
    \begin{cases}
        \pi^2 \nmax^2 \, , & \nmax \ll Q \\
        Q^2 / (2 \chi) \, , & \nmax \gtrsim Q
    \end{cases} \, .
\label{eq:totalEnergy}
\end{equation}
This is schematically illustrated in \cref{fig:powerOutput}, where we show an example of the maximum stored energy $\wtot$, derived from the mode function dynamics, as a function of the quality factor $Q$.
We confirm that there is a critical value of the quality factor beyond which the system saturates and the deposited energy becomes independent of $Q$.

In a more realistic scenario, the total energy deposited inside the cavity governs the power output picked up by an antenna.
Naively, the power output measured in one resonant cycle, $\delta t = 2\pi/\omega_n$, can be at most
\begin{equation}
    P_{\mathrm{out}} \approx -\frac{\delta W}{\delta t} \lesssim \frac{\omega_n}{Q} \wtot \, .
\end{equation}
This means that a large quality factor is an experimental disadvantage if the GW cannot fully excite the resonant cavity mode, since the signal output power would be proportional to $\nmax^2 / Q$, as opposed to $Q$.
Instead, the power output should be integrated over $Q$ cycles to recover the energy deposited by the incoming GW. 
In summary, we find that the quality factor of a cavity has a major impact on its experimental sensitivity, as expected.
However, if the cavity's power output corresponds to the measured signal, increasing the quality factor indefinitely, $Q \to \infty$, will not increase the experimental sensitivity beyond a certain maximum.

\subsection{Primordial black hole mergers}

In general, black hole mergers naturally produce transient GW signals with a time-dependent frequency.
Given that electromagnetic cavities are designed to operate at GHz frequencies, here, we focus on the inspiral phase of PBH mergers.
In this scenario, the maximum GW frequency emitted by the system is governed by the innermost stable circular orbit before the merger begins, with a frequency of $\omega_{\mathrm{ISCO}} = 1.38 \times 10^4 \, \mathrm{Hz} \, (M_{\odot} / m_{\mathrm{PBH}})$~\cite{Ajith:2007kx}.
Therefore, GWs in the GHz range are only produced for PBH masses $m_{\mathrm{PBH}} \lesssim 10^{-5} M_{\odot}$.
The GW strain amplitude corresponding to a PBH binary system depends on the masses involved and the distance $r$ from the binary (see, e.g.,~\cite{Maggiore:2007ulw}),
\begin{equation}
    h_{\omega_n} \approx 1.29 \times 10^{-22} \left( \frac{1 \, \mathrm{kpc}}{r} \right) \left(\frac{m_{\mathrm{PBH}}}{10^{-5} M_{\odot}}\right)^{5/3} \left(\frac{\omega_n}{1 \, \mathrm{GHz}}\right)^{2/3} \, .
\end{equation}
Furthermore, at low frequencies, the change in GW frequency is given by~\cite{Blanchet:1995ez, Maggiore:2007ulw, Aggarwal:2025noe}
\begin{equation}
    \dot{\omega} = \frac{12}{5} 2^{1/3} \left(G \mathcal{M}\right)^{5/3} \omega^{11/3} 
    \approx 3.4 \times 10^{9} \, {\mathrm{Hz}}^2 \left( \frac{m_{\mathrm{PBH}}}{10^{-9} M_\odot} \right)^{5/3} \left( \frac{\omega}{{\mathrm{GHz}}}\right)^{11/3}
    \, ,
\label{eq:PBHFrequencyChange}
\end{equation}
where $G$ denotes the gravitational constant and $\mathcal{M}$ is the so-called chirp mass,
\begin{equation}
    \mathcal{M} = \frac{\left( m_1 m_2 \right)^{3/5}}{\left( m_1 + m_2 \right)^{1/5}} \, .
\end{equation}
For the last approximation we have used $m_1 = m_2 = m_{\mathrm{PBH}}$.
Solving \cref{eq:PBHFrequencyChange}, we find that it can be well approximated by a linear expansion about the time $t_n$ where the GW frequency coincides with a resonance frequency $\omega_n$ of the cavity,
\begin{equation}
    \omega (t) = \omega_n + \dot{\omega}(t-t_n) \, .
    \label{eq:FrequencyLinear}
\end{equation}
We can thus identify the scanning rate as $\chi = \dot{\omega}$ in \cref{eq:omega_of_t_1}.

\begin{figure}[t]
    \centering
    \includegraphics{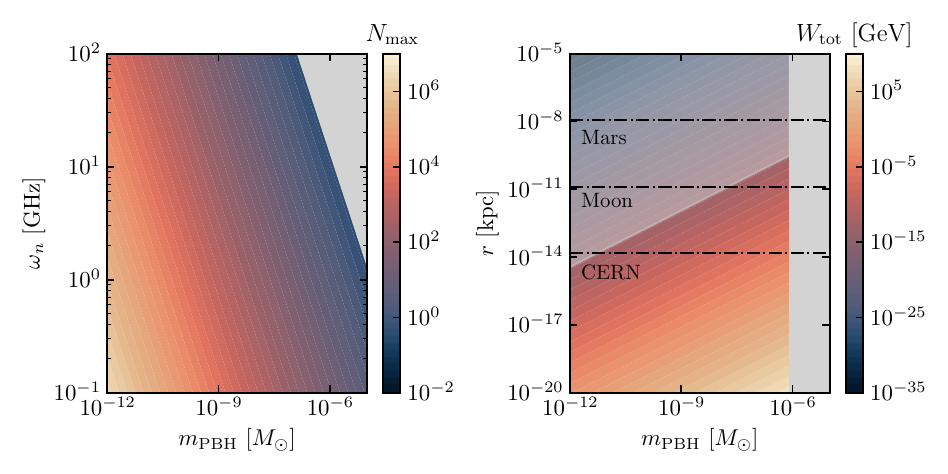}
    \caption{Number of oscillation cycles after which coherence with the resonant oscillations of the cavity is lost, as a function of PBH mass $m_{\mathrm{PBH}}$ and resonance frequency $\omega_n$ (left), and total energy deposited in the cavity by GWs emitted from a PBH merger involving masses $m_{\mathrm{PBH}}$ and at a distance $r$ (right). In both panels, the opaque gray region illustrates the maximum GW frequency that can be emitted by the PBH merger, given by $\omega_{\mathrm{ISCO}}$, and is therefore not part of the viable parameter space. In the right panel, the light gray region indicates energies below the photon energy corresponding to the lowest-lying resonant frequency at $\omega_{011}^{\mathrm{TM}} \approx 16.45 \, \mathrm{GHz}$ (for a spherical cavity of radius $R = 5 \, \mathrm{cm}$). For comparison, the black dash-dotted lines illustrate the mean distance between Mainz and Mars, Mainz and the Moon, and Mainz and CERN.}
    \label{fig:pbh_results}
\end{figure}

Now consider our experimental setup.
The resonance frequencies are fixed by the physical size of the cavity, $\omega_n \sim \vcav^{-1/3}$, and assumed to be in the GHz range.
On the one hand, lighter PBH mergers will remain in the resonance band for a longer period of time.
On the other hand, they will also have a smaller signal amplitude.
As we have mentioned earlier, the dimensionless change in frequency $\dot{\omega} / \omega_n^2$ determines the number of oscillation cycles $\nmax$ for which the GW signal stays in resonance with a given cavity mode, $\nmax \simeq \sqrt{\omega_n^2 / (4\pi \dot{\omega})}$.
The total energy $\wtot$ in \cref{eq:totalEnergy} that can at most be deposited inside the cavity volume is limited by $\nmax$, otherwise is limited by the quality factor $Q$.

We illustrate an example of this in \cref{fig:pbh_results}, where we show the maximum number of resonant cycles $\nmax$ along with the total energy $\wtot$ deposited in the cavity by an incoming GW as a function of the PBH mass and distance of the merger.
Here, we choose a spherical cavity of radius $R=5 \, \mathrm{cm}$ in an external magnetic field with $B_0 = 14 \, \mathrm{T}$, and consider the lowest-lying TM resonance at $\omega_{11}^{\mathrm{TM}} \approx 16.45 \, \mathrm{GHz}$ (see \cref{tab:resfrequenciesSpherical}).
Furthermore, we choose a coupling of the incoming GW to this mode of $\eta_n = 0.3$.
We find that, for signals in viable regions of parameter space, a high quality factor with $Q \gtrsim 10^5$ is rarely beneficial to maximize the experimental sensitivity.
Instead, it is limited by the maximum number of resonant oscillations, $\nmax$, before decoherence sets in.
Therefore, the estimated energy deposited in the cavity is below the TeV range, even for mergers that happen in immediate vicinity of Earth.
See also~\cite{Barrau:2023kuv,Barrau:2024kcb} for related studies.\footnote{Similar calculations will become available in~\cite{GravNetCDR}.}

Clearly, these naive estimates leave room for some improvement of the sensitivity.
In principle, the PBH merger will not only excite the lowest-lying mode, but instead evolve through several more resonance bands of the cavity, which increases the total energy deposited.
Similarly, the combination of a large number of cavities (see, e.g., \cite{Schmieden:2023fzn, Schneemann:2024qli}), or even using one very large cavity, would increase the deposited energy significantly.
However, this assumes that, at the same time, the noise can be suppressed sufficiently.
We also remark that for $\nmax \lesssim 1$, the linear time evolution for the scanning frequency in \cref{eq:FrequencyLinear} becomes inaccurate.
Unfortunately, this is precisely the region of parameter space where one would expect the largest signal.

\section{Conclusions}
\label{sec:conclusions}

In addition to axion searches, electromagnetic cavity experiments may be used to probe the conversion of GWs into photons in the vicinity of an external magnetic field.
Their characteristic size allows to target GW signals in the GHz range and beyond.

In this work, we have estimated the experimental sensitivity of cylindrical as well as spherical cavities to both monochromatic and transient GW signals.
For simplicity, we have focused on GW frequencies close to the electromagnetic resonances of the cavity.
In this case, their sensitivity is dominated by the total energy stored in the oscillations of the electromagnetic fields inside the cavity, which is deposited by the incoming GW.
For monochromatic GWs, this energy and its associated power output can be captured by a dimensionless coupling coefficient which parametrizes the overlap between the effective current induced by the incoming GW and the electromagnetic resonances.
Although this inherently neglects any detector motion perceived by the observer, it serves as a robust approximation of the expected signal strength when expressed in TT gauge~\cite{Ratzinger:2024spd}.
That said, we have calculated the GW couplings to the lowest-lying resonant modes and find that, for both the cylindrical and the spherical cavity, they strongly depend on the incidence angle of the incoming GW relative to the external magnetic field.
The couplings are typically maximal, reaching values up to $\eta \approx 0.4$, if the GW is propagating perpendicular to the magnetic field, whereas they diminish to zero for GWs traveling (anti-)parallel to it.
This suggests that a practical experimental scenario could benefit from the integration of multiple cavities, each oriented with a distinct alignment of the external magnetic field, similar to the concepts proposed in GravNet~\cite{Schmieden:2023fzn}.

While for monochromatic GWs close to the resonant frequency the sensitivity is primarily governed by the associated coupling coefficient, any nontrivial dynamics of the GW signal may drastically alter this estimate.
As an example, we have investigated transient signals from PBH mergers.
At low frequencies, their emitted GW frequency increases linearly with time.
The signal driven by the effective current therefore traverses the different resonant frequency bands within the cavity.
By solving the relevant equations of motion for the electromagnetic modes, we observe that their resonant amplification is limited either by a finite quality factor, or by a maximum number of oscillations inside the cavity before decoherence occurs.
Likewise, the total energy is limited by the maximum resonant amplification.
We observe that, for transient signals, a high quality factor with $Q \gtrsim 10^5$ is rarely beneficial to maximize the experimental sensitivity.
Even in the most optimistic scenario, only PBH mergers occurring within the solar system deposit an observable amount of energy inside the cavity, presenting a significant challenge for their detection.

Nevertheless, in the future, the experimental sensitivity of electromagnetic cavities to PBH mergers may be enhanced by recognizing that these events do not only deposit energy in a single resonance band of the cavity.
Instead, it may subsequently excite \emph{all} other resonant modes, as the frequency continuously changes.
In addition, combining multiple cavity experiments may lead to a significant increase in the total deposited energy.
To fully understand and unlock the potential of electromagnetic cavity experiments in the search for GWs in the high-frequency regime, these options certainly merit further investigation.

\section*{Acknowledgments}

We would like to thank W.~Ratzinger, T.~Schneemann and J.~Sch\"utte-Engel for useful discussions, and M.~Fritz, M.~Schott and M.~Wurm for collaboration at initial stages of this project.
The authors are supported by the Cluster of Excellence \emph{Precision Physics, Fundamental Interactions and Structure of Matter} (PRISMA$^+$ EXC 2118/1) funded by the German Research Foundation (DFG) within the German Excellence Strategy (Project No. 390831469).

\appendix

\section{Electromagnetic Resonances of Cavities}
\label{app:modesCavity}

For incoming GWs with a wave length which is of order of the characteristic length scale of the electromagnetic cavity, $\omega R \gtrsim 1$, the electromagnetic resonances of the experiment dominate the signal.
These resonances depend on the cavity's geometry and are eigenfunctions of the Helmholtz operator with appropriate boundary conditions, schematically written as~\cite{hill2009electromagnetic}
\begin{equation}
    \left( \nabla^2 + \omega^2 \right) \psi = 0 \, .
\end{equation}
The resonant modes enter the dimensionless coupling coefficients $\eta$ in \cref{eq:couplingCoefficient}.
For completeness, we briefly present them here for both a cylindrical and a spherical cavity geometry.
We closely follow~\cite{hill2009electromagnetic}, where a detailed derivation and discussion can be found.

\subsection{Cylindrical cavities}
\label{app:modesCylindricalCavity}

We consider a cylindrical cavity of radius $R$ and total length $L$, and choose a coordinate system where the cylinder's symmetry axis aligns with the $z$-axis, extending from $z=0$ to $z=L$.
The electromagnetic resonances of the cavity are governed by three quantum numbers, the azimuthal $n$, the radial $p$, and the longitudinal quantum number $q$.
For any $n \neq 0$, each mode is doubly degenerate.
The resonant frequencies of the transverse magnetic (TM) and transverse electric (TE) are given by~\cite{hill2009electromagnetic}
\begin{align}
    \omega_{npq}^{\mathrm{TM}} &= \sqrt{\left( \frac{x_{np}}{R} \right)^2 + \left( \frac{q \pi}{L} \right)^2} \, , \\
    \omega_{npq}^{\mathrm{TE}} &= \sqrt{\left( \frac{x_{np}^{\prime}}{R} \right)^2 + \left( \frac{q \pi}{L} \right)^2} \, .
\end{align}
Here, $x_{np}$ denotes the $p$-th root of the $n$-th order Bessel function of the first kind, $J_n(x_{np}) = 0$, while $x_{np}^{\prime}$ is the $p$-th root of its derivative, $J_n^{\prime}(x_{np}^{\prime}) = 0$.

In cylindrical coordinates, the TM mode functions read~\cite{hill2009electromagnetic}
\begin{align}
    E_{\rho, npq}^{\mathrm{TM}} &= -\frac{E_0}{\omega_{npq}^2 - \left(\frac{q \pi}{L}\right)^2} \frac{q \pi}{L} \frac{x_{np}}{R} J_n^{\prime} \left(\frac{x_{np}}{R} \rho \right) \sin \left(\frac{q \pi}{L} z \right) \begin{Bmatrix}
        \sin \left(n \phi\right) \\
        \cos \left(n \phi\right)
    \end{Bmatrix} \, , \\
    E_{\phi, npq}^{\mathrm{TM}} &= -\frac{E_0}{\omega_{npq}^2 - \left(\frac{q \pi}{L}\right)^2} \frac{q \pi}{L} \frac{n}{\rho} J_n \left(\frac{x_{np}}{R} \rho \right) \sin \left(\frac{q \pi}{L} z \right) \begin{Bmatrix}
        \cos \left(n \phi\right) \\
        -\sin \left(n \phi\right)
    \end{Bmatrix} \, , \\
    E_{z, npq}^{\mathrm{TM}} &= E_0 J_n \left(\frac{x_{np}}{R} \rho \right) \cos \left(\frac{q \pi}{L} z \right) \begin{Bmatrix}
        \sin \left(n \phi\right) \\
        \cos \left(n \phi\right)
    \end{Bmatrix} \, ,
\end{align}
while the corresponding TE modes are given by~\cite{hill2009electromagnetic}
\begin{align}
    E_{\rho, npq}^{\mathrm{TE}} &= i E_0 \frac{\omega_{npq}}{\omega_{npq}^2 - \left(\frac{q \pi}{L}\right)^2} \frac{n}{\rho} J_n \left(\frac{x_{np}^{\prime}}{R} \rho \right) \sin \left(\frac{q \pi}{L} z \right) \begin{Bmatrix}
        \cos \left(n \phi\right) \\
        -\sin \left(n \phi\right)
    \end{Bmatrix} \, , \\
    E_{\phi, npq}^{\mathrm{TE}} &= -i E_0 \frac{\omega_{npq}}{\omega_{npq}^2 - \left(\frac{q \pi}{L}\right)^2} \frac{x_{np}^{\prime}}{R} J_n^{\prime} \left(\frac{x_{np}^{\prime}}{R} \rho \right) \sin \left(\frac{q \pi}{L} z \right) \begin{Bmatrix}
        \sin \left(n \phi\right) \\
        \cos \left(n \phi\right)
    \end{Bmatrix} \, , \\
    E_{z, npq}^{\mathrm{TE}} &= 0 \, .
\end{align}
Note that, here, we have introduced the arbitrary normalization factor $E_0$, which could be in principle be different for each mode.
For our purposes, however, this normalization does not alter the computation of the coupling coefficients $\eta$ in \cref{eq:couplingCoefficient}.

\subsection{Spherical cavities}
\label{app:modesSphericalCavity}

We now consider a spherical cavity of radius $R$ centered at the origin of our coordinate system.
Similar to the cylinder, the electromagnetic resonances of the spherical cavity are governed by three quantum numbers, the azimuthal $m$, the radial $p$, and the longitudinal quantum number $n$.
The resonant frequencies are given by~\cite{hill2009electromagnetic}
\begin{align}
    \omega_{mnp}^{\mathrm{TM}} &= \frac{u_{np}^{\prime}}{R} \, , \\
    \omega_{mnp}^{\mathrm{TE}} &= \frac{u_{np}}{R} \, .
\end{align}
Here, $u_{np}$ is the $p$-th root of the $n$-th order spherical Bessel function of the first kind, $j_n (u_{np}) = 0$, and $u_{np}^{\prime}$ is the $p$-th root of the derivative of the $n$-th order Harrington's spherical Bessel function, $\Jhat_n^{\prime} (u_{np}^{\prime}) = 0$.
The latter is defined as
\begin{equation}
    \Jhat_n (x) = x j_n(x) \, .
\end{equation}
Clearly, the resonant frequencies do not depend on the azimuthal quantum number.
Therefore, similar to the quantization of the hydrogen atom, each mode is $(2p+1)$-times degenerate, with $m = -p, \ldots, p$.
An example of resonant frequencies for a cavity with radius $R=5 \, \mathrm{cm}$ is illustrated in \cref{tab:resfrequenciesSpherical}.

\begin{table}[t]
    \centering
    \begin{tabular}{rcccccccc}
    \toprule
                & \multicolumn{4}{c}{$\omega_{mnp}^{\mathrm{TE}}$ [GHz]} & \multicolumn{4}{c}{$\omega_{mnp}^{\mathrm{TM}}$ [GHz]} \\
                \cmidrule(lr){2-5} \cmidrule(lr){6-9}
        $n/p$   &       1 &       2 &       3 &         4   &
                        1 &       2 &       3 &         4   \\
    \midrule
            1   &   26.94 &   46.32 &   65.38 &     84.34   &
                    16.45 &   36.68 &   55.86 &     74.87   \\
            2   &   34.56 &   54.53 &   73.89 &     93.03   &
                    23.21 &   44.63 &   64.24 &     83.47   \\
            3   &   41.90 &   62.46 &   82.13 &    101.48   &
                    29.82 &   52.30 &   72.33 &     91.82   \\
    \bottomrule
    \end{tabular}
    \caption{Resonant frequencies $\omega_{mnp}$ of the TE (left) and TM (right) modes of a spherical cavity with radius $R=5 \, \mathrm{cm}$.
    The rows illustrate different values of $n$, while the columns show different values of $p$. The azimuthal quantum number can take values $m = -p , \ldots , p$.}
    \label{tab:resfrequenciesSpherical}
\end{table}

Adopting spherical coordinates, the TM mode functions read~\cite{hill2009electromagnetic}
\begin{align}
    E_{\rho, mnp}^{\mathrm{TM}} &= i \frac{n(n+1)}{\omega_{mnp}^2 \rho^2} \Jhat_n \left( \frac{u_{np}^{\prime}}{R} \rho \right) P_n^m (\cos\theta) \begin{Bmatrix}
        E_+ \cos \left(m \phi\right) \\
        E_- \sin \left(m \phi\right)
    \end{Bmatrix} \, , \\
    E_{\theta, mnp}^{\mathrm{TM}} &= \frac{i}{\omega_{mnp} \rho} \Jhat_n^{\prime} \left( \frac{u_{np}^{\prime}}{R} \rho \right) \frac{\md}{\md \theta} P_n^m (\cos\theta) \begin{Bmatrix}
        E_+ \cos \left(m \phi\right) \\
        E_- \sin \left(m \phi\right)
    \end{Bmatrix} \, , \\
    E_{\phi, mnp}^{\mathrm{TM}} &= \frac{i m}{\omega_{mnp} \rho \sin \theta} \Jhat_n^{\prime} \left( \frac{u_{np}^{\prime}}{R} \rho \right) P_n^m (\cos\theta) \begin{Bmatrix}
        -E_+ \sin \left(m \phi\right) \\
        E_- \cos \left(m \phi\right)
    \end{Bmatrix} \, ,
\end{align}
while the corresponding TE modes are given by~\cite{hill2009electromagnetic}
\begin{align}
    E_{\rho, mnp}^{\mathrm{TE}} &= 0 \, , \\
    E_{\theta, mnp}^{\mathrm{TE}} &= \frac{m}{\omega_{mnp} \rho \sin \theta} \Jhat_n \left( \frac{u_{np}}{R} \rho \right) P_n^m (\cos\theta) \begin{Bmatrix}
        E_+ \sin \left(m \phi\right) \\
        -E_- \cos \left(m \phi\right)
    \end{Bmatrix} \, , \\
    E_{\phi, mnp}^{\mathrm{TE}} &= \frac{1}{\omega_{mnp} \rho} \Jhat_n \left( \frac{u_{np}}{R} \rho \right) \frac{\md}{\md \theta} P_n^m (\cos\theta) \begin{Bmatrix}
        E_+ \cos \left(m \phi\right) \\
        E_- \sin \left(m \phi\right)
    \end{Bmatrix} \, .
\end{align}
Here, we have introduced the arbitrary normalization factor $E_{\pm}$, which could be in principle be different for each mode.
For our purposes, however, this normalization does not alter the computation of the coupling coefficients $\eta$ in \cref{eq:couplingCoefficient}.

\section{Coupling Coefficients in Fermi-Normal Coordinates}
\label{app:fnCoordinates}

\begin{figure}[t]
	\centering
    \includegraphics{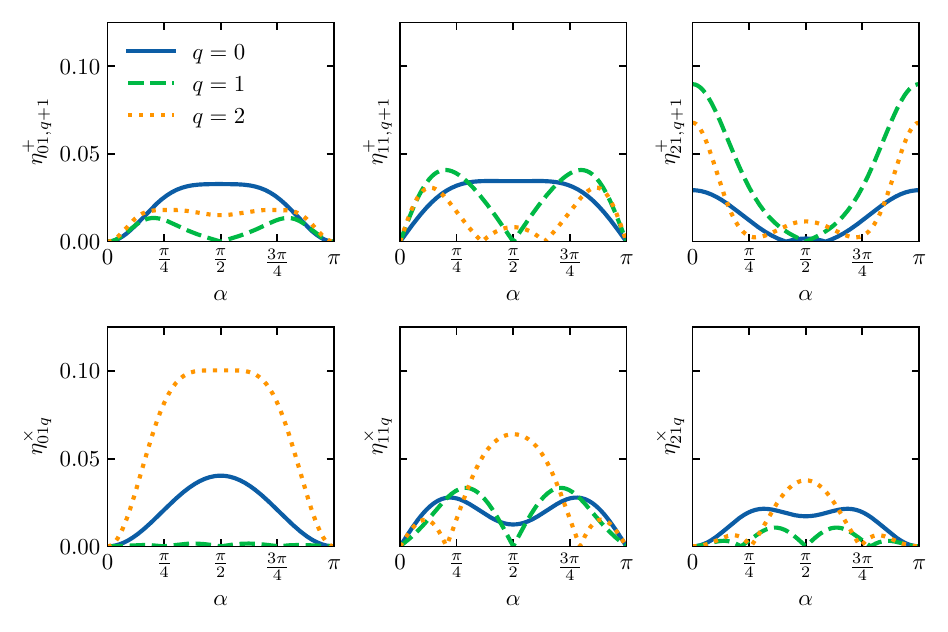}
	\caption{Coupling coefficients of a cylindrical cavity, similar to \cref{fig:etasCylindrical}, but parametrized in FN coordinates. Note that the normalization of the effective current induced by the incoming GW differs by an additional factor of $\omega \vcav^{1/3}$, such that the comparison of sensitivities between TT gauge and FN coordinates has to be taken with care (see main text).}
	\label{fig:etasCylindricalFN}
\end{figure}

In general, the electromagnetic response of a cavity to an incoming GW is independent of the chosen coordinate frame.
In addition to the electromagnetic excitations, this requires a complete implementation of mechanical detector deformations as well as the observer's perceived motion~\cite{Ratzinger:2024spd}.
In other words, the mechanical and electromagnetic properties of the cavity and the sensor have to be carefully characterized.
Nevertheless, in some cases, choosing a particular coordinate frame and neglecting the detector's motion may still provide a viable approximation to the exact signal.

Close to the (electromagnetic) resonant frequencies of the cavity, the electromagnetic excitations are parametrically enhanced with respect to the detector's mechanical response.
This suggests that, in this regime, the latter can be neglected to approximate the measured signal, and capture it by a coupling coefficient $\eta$ associated to each mode instead.
Typical approximation schemes include a free-falling or a rigid approximation, where the detector motion is neglected either in TT gauge or FN coordinates.
These are suitable for capturing the regimes where $\omega L \gg v_s$ and $\omega L \ll v_s$, respectively~\cite{Ratzinger:2024spd}, where $L$ is the characteristic length scale of the cavity and $v_s$ is the velocity of sound inside the detector's material, which is typically of the order $v_s \sim \mathcal{O}(10^{-5})$.
Therefore, the free-falling approximation, using TT gauge, is the optimal choice for scenarios where the GW frequency is of the same order as the size of the experiment, $\omega L \sim 1$.
We show the corresponding coupling coefficients in \cref{fig:etasCylindrical,fig:etasSpherical}.

\begin{figure}[t]
	\centering
    \includegraphics{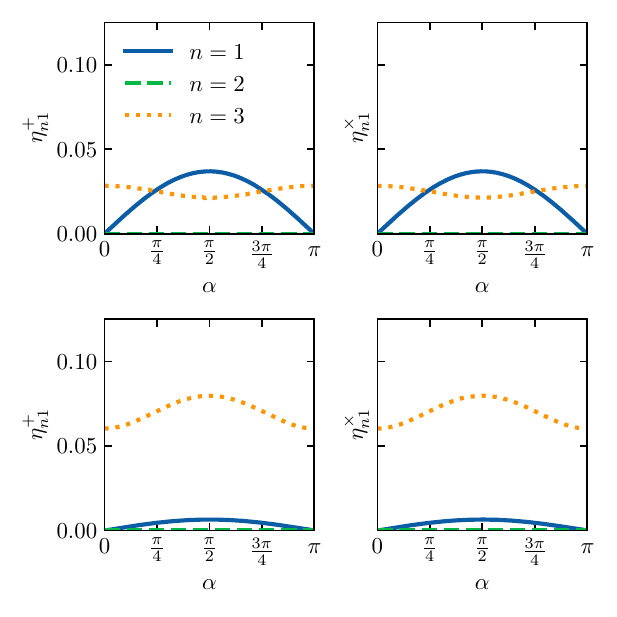}
	\caption{Coupling coefficients of a spherical cavity, similar to \cref{fig:etasSpherical}, but parametrized in FN coordinates. Note that the normalization of the effective current induced by the incoming GW differs by an additional factor of $\omega \vcav^{1/3}$, such that the comparison of sensitivities between TT gauge and FN coordinates has to be taken with care (see main text).}
	\label{fig:etasSphericalFN}
\end{figure}

These coupling coefficients, and the associated expected experimental sensitivity, can drastically differ for other coordinate frames.
For comparison, we illustrate the coupling coefficients in FN coordinates for a cylindrical cavity in \cref{fig:etasCylindricalFN} and for a spherical cavity in \cref{fig:etasSphericalFN}.
For the cylindrical cavity, these reproduce the results originally found in~\cite{Berlin:2021txa}, when taking into account an additional factor $\sqrt{2}$ due to the different normalization of the metric tensor.
Naively, for both the cylindrical and the spherical cavity, their overall magnitude appears smaller compared to the couplings obtained in TT gauge.
However, we remark that the normalization of the effective current induced by the incoming GW differs by an additional factor of $\omega_g \vcav^{1/3}$ (see~\cite{Berlin:2021txa}), such that, in this sense, the normalization of $\eta$ in both scenarios is not the same.
Hence, the comparison of sensitivities between TT gauge and FN coordinates has to be taken with care.
More importantly, some modes suggest an enhanced sensitivity for incoming GWs propagating (anti-)parallel to the magnetic field in FN coordinates, while the sensitivity of this scenario vanishes entirely in TT gauge.
This illustrates the drastic frame dependence of the experimental sensitivity parametrized in terms of a dimensionless coupling coefficient.

\section{The Driven Harmonic Oscillator}
\label{app:DrivenHarmonicOscillator}

This section is aimed at collecting a few key properties of the damped harmonic oscillator, driven by an external force.
\cref{eq:eomModeFunctions} describes the dynamics of a driven harmonic oscillator, where the driving force has a time-dependent frequency.
It is straightforward to solve these equations of motion using Green's functions, which satisfy
\begin{equation}
    \left( \frac{\md^2}{\md t^2} + 2 \gamma \frac{\md}{\md t} + \omega_0^2 \right) G(t, t^{\prime}) = \delta(t-t^{\prime}) \, ,
\end{equation}
where in our case the second term characterizes the damping due to the quality factor of the cavity, $2 \gamma = \omega_0 / Q$.
An underdamped system, $\gamma < \omega_0$ has the solution (see, e.g., \cite{wess2009mechanics})
\begin{equation}
    G(t,t^{\prime}) = \Theta(t-t^{\prime}) \frac{\me^{-\gamma(t-t^{\prime})}}{\sqrt{\omega_0^2 - \gamma^2}} \sin \left[ \sqrt{\omega_0^2 - \gamma^2}(t-t^{\prime})\right] \, .
\label{eq:GreensFunctionHO}
\end{equation}
Given a time-dependent driving force $f(t)$, the general solution to the equations of motion is of the form
\begin{equation}
    x(t) = \int_{-\infty}^{\infty} \md t^{\prime} \, G(t,t^{\prime}) f(t^{\prime}) + \ldots \, ,
\end{equation}
where the dots denote terms that depend on the initial conditions, which we can neglect.

\begin{figure}[t]
    \centering
    \includegraphics{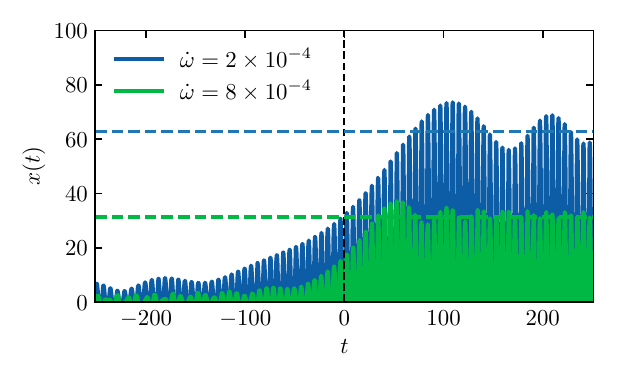}
    \caption{Approximate solutions of the driven harmonic oscillator for negligible damping $Q \to \infty$. For simplicity, we choose $\omega_0=1$, and a driving frequency $\omega(t) = \omega_0 + \dot{\omega} t$, such that the resonance frequency is matched at $t=0$. The different colors illustrate the different frequency scanning rates $\dot{\omega}$, while the colored dashed lines show an estimate for the corresponding maximum amplitude given in \cref{eq:AmplitudeMaximumEstimate}.}
    \label{fig:HOexcited}
\end{figure}

In general, we are interested in a periodic driving force with a frequency that increases with time.
For simplicity, let us consider a force that is turned on instantaneously, at $t=0$, with a fixed amplitude,
\begin{equation}
    f(t) = \Theta(t) \sin \left[ \omega_0 \left(1 + \frac{\dot{\omega}}{\omega_0} t \right) t\right] \, .
\label{eq:HODrivingForce}
\end{equation}
Let us now approximate the dynamical behavior of $x(t)$ excited by this driving force.
The time scale for a single oscillation is $t_{\mathrm{osc}} = 2\pi / \omega_0$, while the exponential term of the Green's function~\eqref{eq:GreensFunctionHO} implies that the damping is only relevant on time scales $t_{\mathrm{damp}} \sim 1 / \gamma = 2Q / \omega_0$.
In other words, for a cavity with a high quality factor, the time scale of damping is negligible compared to the time scale where a coherent state of photons develops.
In the limit of weak damping, $Q \to \infty$, the solution asymptotically approaches
\begin{equation}
    x(t) \sim \omega_0^{-1} \int_{-\infty}^{\infty} \md t^{\prime} \, \Theta(t-t^{\prime}) \sin \left[ \omega_0 (t-t^{\prime}) \right] \sin \left[ \omega_0 \left(1 + \frac{\dot{\omega}}{\omega_0} t^{\prime} \right) t^{\prime} \right] \, .
\label{eq:SolutionHarmonicOscillatorApproximation}
\end{equation}
The amplitude $A$ of the oscillations grows as long as both sine-functions are oscillating in phase.
In this case, during one oscillation it roughly grows by an amount $\delta A$, given by
\begin{equation}
    \delta A \simeq \omega_0^{-1} \int_{0}^{t_{\mathrm{osc}}} \md t^{\prime} \, \sin^2 (\omega_0 t^{\prime}) = \frac{\pi}{\omega_0^2} \, .
\end{equation}
We again remark that this estimate holds in the asymptotic limit of negligible damping, $Q \to \infty$.
Naively, a resonant growth will occur until the phases of \cref{eq:SolutionHarmonicOscillatorApproximation} are shifted by a factor $\pi$, such that they contribute with opposite sign.
Beyond this time scale, the amplitude decreases again, and $x(t)$ eventually begins to oscillate close to its maximum value.
Therefore, there is a maximum number of oscillations $\nmax$ during which the amplitude grows, before decoherence occurs.
From \cref{eq:SolutionHarmonicOscillatorApproximation} we can estimate this time scale as
\begin{equation}
    t_{\mathrm{decoherence}} \simeq \sqrt{\frac{\pi}{\dot{\omega}}} \, .
\end{equation}
Using that $t_{\mathrm{decoherence}} = \nmax t_{\mathrm{osc}}$, we can estimate the maximum number of oscillations during which the amplitude grows,
\begin{equation}
    \nmax \simeq \sqrt{\frac{\omega_0^2}{4\pi \dot{\omega}}} \, .
\end{equation}
In the limit of negligible damping, $Q \to \infty$, we therefore find the maximum amplitude of oscillations to be approximated by
\begin{equation}
    A_{\mathrm{max}} \simeq \nmax \delta A \simeq \sqrt{\frac{\pi}{4 \omega_0^2 \dot{\omega}}} \, .
\label{eq:AmplitudeMaximumEstimate}
\end{equation}
As we have mentioned earlier, the damping is only relevant on time scales $t_{\mathrm{damp}} \simeq 2Q/\omega_0$, while the time scale after which the oscillations have reached their maximum amplitude is $t_{\mathrm{max}} \simeq 2 \pi \nmax / \omega_0$.
Therefore, the damping can be neglected as long as $Q \gg \pi \nmax$.

In \cref{fig:HOexcited}, we show an example for the approximate solution $x(t)$ given in \cref{eq:SolutionHarmonicOscillatorApproximation} for two different values of the frequency-scanning rate $\dot{\omega}$, along with our estimate for the maximum amplitude.
Note that, here, the driving force scans through the resonance frequency from the infinite past, i.e.~we drop the $\Theta(t-t^{\prime})$ term in \cref{eq:HODrivingForce}.
In this case, the time scale of coherence and the corresponding resonant growth of the amplitude is somewhat increased, such that the maximum amplitude slightly exceeds our naive estimates.

Going back to \cref{eq:cavity_oscillator}, we find that the driving force is proportional to the derivative of the effective current.
This, and the overall scaling of the current with frequency, will yield additional powers of the resonance frequency $\omega_0$ in the equations of motion.
Furthermore, the time-dependence of the GW frequency is not linear, and also the amplitude depends on time.
These effects are negligible given that the resonances are narrow, but will play a role when comparing the power deposited into different resonant cavity modes.

\bibliographystyle{inspire}
\bibliography{refs, refs_noninspire}

\end{document}